\documentclass[12pt]{article}
\usepackage{amsmath}
\usepackage{epsfig}
\usepackage {graphicx}
\def\email#1{\date{\tt#1}}
\def\address#1{\par\bigskip\noindent\small#1}
\def\balpha{\mbox {\boldmath $\alpha$} }

\def\btau{\mbox {\boldmath $\tau$} }

\def\can{c^A_{pj}({\bf n})}
\def\cb{c^{B(k)}_{pj}}
\def\vcb{{\bf c}^{B(k)}_{pj}}
\def\cbn{c^{B(k)}_{pj}({\bf n})}
\def\vcan{{\bf c}^A_{pj}({\bf n})}
\def\vLb{{\bf L}^{B(k)}_j}
\def\vcgb{{\bf c}^{B(k)}_{gj}}
\def\vcgan{{\bf c}^A_{gj}({\bf n})}

\begin{document}

\title{FUNDAMENTAL SOLUTIONS IN PLANE PROBLEM FOR ANISOTROPIC ELASTIC MEDIUM UNDER MOVING OSCILLATING SOURCE}
\author{G.~Iovane  \and A.V.~Nasedkin \and M.~Ciarletta}

% IF YOU HAVE NO E-MAIL PLEASE INSERT EMPTY ARGUMENT: \email{}
\email{iovane@diima.unisa.it, nasedkin@math.rsu.ru,
ciarlett@diima.unisa.it}

\maketitle \thispagestyle{empty}

\begin{abstract}
In present article we consider the problems of concentrated point
force which is moving with constant velocity and oscillating with
cyclic frequency in unbounded homogeneous anisotropic elastic
two-dimensional medium. The properties of plane waves and their
phase, slowness and ray or group velocity curves for 2D problem in
moving coordinate system are described. By using the Fourier
integral transform techniques and established the properties of
the plane waves, the explicit representation of the elastodynamic
Green's tensor is obtained for all types of source motion as a sum
of the integrals over the finite interval. The dynamic components
of the Green's tensor are extracted.

The stationary phase method is applied to derive an asymptotic
approximation of the far wave field. The simple formulae for
Poynting energy flux vectors for moving and fixed observers are
presented too. It is noted that in the far zones the cylindrical
waves are separated under kinematics and energy.

It is shown that the motion bring some differences in the far
field properties. They are modification of the wave propagation
zones and their number, fast and slow waves appearance under
trans- and superseismic motion and so on.
\end{abstract}

\section{Introduction}

The theory of elasticity with moving source have a range of
important applications. They are linked with the developing of
more high-speed transport and with the necessity to evaluate the
influence of elastic waves from moving objects on different
constructions. The theoretical side of these problems is also of
interest. The changes in the character of mechanical fields is
essentially depend on the behavior of source motion and are caused
by the type of differential equations ranging from elliptical to
hyperbolic.

From the steady-state problems with moving sources the problems
with moving and oscillating sources are the most complicated. We
use the following terminology. If the source is moving with
constant velocity ${\bf w}$ and simultaneously oscillating with
frequency $\omega$, we consider the problem $B$, while in the case
${\bf w}=0$, we have the classical harmonic problems with
oscillating source and so refer them to the problem $A$.

Nowadays considerable progress is obtained in the investigation of
the problem $B$. The correspondence principles between the
problems $B$ and $A$ [1], analogous known in fluid mechanic [2],
are stated. The principles for unique solution selection are
investigated [1, 3]. The energetic principles are formulated and
the general theorems about energy transport are established [3],
the some actual problems $B$ for isotropic media are solved
[4--6].

It is evident, that for anisotropic elastic media the particular
problems $B$ are not studied sufficiently. In [7] the plane waves
and fundamental solutions for problem $B$ for tree-dimensional
anisotropic media are studied, and in [8] the approaches for the
problem $B$ for anisotropic elastic and piezoelectric waveguides
are proposed.

In present paper we investigate the properties of plane waves and
fundamental solutions in the problem $B$ for anisotropic elastic
plane.

It is well-known, that for many classes of anisotropic materials
it is possible to formulate the problems $A$ and $B$ in conditions
of plane deformation. For existence of the plane problem several
elastic modules must be equal to zero. Thus, for the plane
deformation in the plane $0x_1x_3$ for elastic modules in
two-index notation we require that
$C_{14}=C_{34}=C_{45}=C_{16}=C_{36}=C_{56}=0$, for the plane
deformation in the plane $0x_1x_2$
$C_{14}=C_{24}=C_{46}=C_{15}=C_{25}=C_{56}=0$, and for the plane
deformation in the plane $0x_2x_3$
$C_{25}=C_{26}=C_{35}=C_{36}=C_{45}=C_{46}=0$.

The vectors of external force in the plane deformation problems
must not have non-zero components in the perpendicular direction
and must not depend on perpendicular to plane coordinate.

We assume that all above-mentioned conditions are implemented. To
be more concrete, let us accept that the plane deformation is
realized in plane $0x_1x_2$, and therefore the vector of mechanic
displacement has non-zero components $u_1$ and $u_2$: ${\bf u} =
\{ u_1({\bf x},t), u_2({\bf x},t)\}$, ${\bf x}=\{ x_1, x_2\}$. For
harmonic vibrations with frequency $\omega$ we shall find the
established solution in the form
$$
{\bf u}={\bf v} {\rm exp} \, (i\omega t).\eqno(1.1)
$$

In the problem with moving sources we consider two coordinate
systems. Let $\{ \xi_1, \xi_2 \}$ be fixed coordinate system with
time $\tau$, and let $\{ x_1, x_2\}$ be coordinate system, which
is moving relatively to fixed system with constant velocity ${\bf
w}=\{ w_1, w_2\}$. We denote $t$ the time in moving coordinate
system. The two coordinate systems are connected with each other
by the following relations
$$
x_1=\xi_1-w_1\tau, \quad x_2=\xi_2-w_2\tau, \quad t=\tau .
\eqno(1.2)
$$

For the problem $B$ we propose that in moving coordinate system
the harmonic behavior (1.1) with frequency $\omega$ is existed.

For the equations of theory of elasticity in moving coordinate
system we use the following relations from (1.2)
$$
\nabla^{\xi}=\nabla^x, \quad \partial_{\tau}=\partial_t-{\bf w}
\cdot \nabla^x, \eqno(1.3)
$$
where $\nabla^x=\{ \partial /\partial x_1, \partial /\partial
x_2\}=\{ \partial_1, \partial_2\}$, $\nabla^{\xi}=\{ \partial
/\partial \xi_1, \partial /\partial \xi_2\}$.

Then the stain vector ${\bf S}=\{\varepsilon_{11},
\varepsilon_{22}, 2\varepsilon_{12}\}$ ($\varepsilon_{ij}$ are the
components of strain tensor) may be expressed from the
displacement vector as
$$
{\bf S}={\bf L}(\nabla^x) \cdot {\bf u}, \eqno(1.4)
$$
where
$$
{\bf L} (\nabla^x)= \left(
\begin{array}{cc}
\partial_1 & 0 \\
0 & \partial_2 \\
\partial_2 & \partial_1
\end{array}
\right). \eqno(1.5)
$$

From Hook's low for anisotropic media for the plane deformation in
the plane ($x_1x_2$) we have the relation between the stress
vector ${\bf T}=\{\sigma_{11}, \sigma_{22}, \sigma_{12}\}$
($\sigma_{ij}$ are the components of stress tensor) and the strain
vector ${\bf S}$
$$
{\bf T}={\bf C}\cdot {\bf S}, \eqno(1.6)
$$
where
$$
{\bf C} = \left(
\begin{array}{ccc}
C_{11} & C_{12} & C_{16} \\
 & C_{22} & C_{26} \\
{\rm sym} &  & C_{66}
\end{array}
\right). \eqno(1.7)
$$

We suppose here $C_{ij}$ is not depended on coordinate, i.e. the
medium material is homogeneous.

By using (1.3) and the previous notations the equations for
elastic medium in the case of plane deformation may be written in
the form
$$
{\bf L}^* (\nabla^x) \cdot {\bf T}+ {\bf f} = \rho (\partial_t -
{\bf w} \cdot \nabla^x)^2 {\bf u}, \eqno(1.8)
$$
where $\rho$ is the density ($\rho = {\rm const}$), ${\bf f}$ is
the body force, $(...)^*$ is the conjugation operator.

By using (1.4)---(1.6) the equation (1.8) may be rewritten
$$
{\bf L}^* (\nabla^x) \cdot {\bf C} \cdot {\bf L} (\nabla^x) \cdot
{\bf u}+ {\bf f} = \rho (\partial_t - {\bf w} \cdot \nabla^x)^2
{\bf u}. \eqno(1.9)
$$

This equation together with the (1.1) is the equation for
amplitude {\bf v}.

We study the properties of plane wave in problem $B$ at first, to
determine the fundamental solution.

%\newpage
\setcounter{equation}{3}
\section{Plane waves and their characteristic curves}

We find the solution of equations (1.9) without body force (${\bf
f}=0$) in the form of plane waves
$$
{\bf u} = A {\bf p} \, {\rm exp} \, [ i (\omega t - \balpha \cdot
{\bf x}) ], \eqno(2.1)
$$
where $A$ is the amplitude, ${\bf p}$ is the unit polarization
vector ($|{\bf p}| = 1$), $\balpha$ is the wave vector
($\balpha=\alpha {\bf n}$, $|{\bf n}|=1$, ${\bf n}$ is the unit
wave normal vector).

By substituting (2.1) into (1.9) yields the eigenvalue problem
$$
{\bf \Gamma}({\bf n}) \cdot {\bf p}=\rho \nu^2({\bf n}){\bf p},
\eqno(2.2)
$$
where
$$
\nu^2({\bf n})=(c^B_p({\bf n})+w_n)^2 , \eqno(2.3)
$$
$w_n={\bf w}\cdot{\bf n}$, $c^B_p({\bf n})=\omega/\alpha$ is the
phase velocity in the problem $B$, ${\bf \Gamma}({\bf n})={\bf
L}^* ({\bf n})\cdot{\bf C}\cdot{\bf L}({\bf n})$ is the acoustic
Christoffel's tensor (matrix).

As it follows from common properties of the problem $A$ in 3D [9,
10], here in the problem $A$ in 2D two plane waves with phase
velocities $\nu_j({\bf n}) = \can$, $\can >0$, $j=1,2$, and with
polarization vector (eigenvector) ${\bf p}_j$ exist for arbitrary
direction ${\bf n}$,which can be selected orthonormalized. For the
problem $B$, as it is obvious from (2.2), the situation is more
complicated. Particularly, for the direct plane waves
($c^B_{pj}>0$) in the problem $B$ with fixed ${\bf n}$ and
$j\in\{1,2\}$
$$
(\can >  w_n) \; \wedge \; (\can \geq  -w_n) \quad \Rightarrow
\quad c^B_{pj}({\bf n})=\can - w_n, \eqno(2.4)
$$
$$
\can < -w_n \quad \Rightarrow \quad c^{Bk}_{pj}({\bf
n})=(-1)^k\can-w_n; \quad k=0,1, \eqno(2.5)
$$
$$
\can \leq w_n  \quad \Rightarrow \quad \mbox{there are not direct
plane waves}. \eqno(2.6)
$$

Thus, in condition (2.4) for fixed ${\bf n}$ and $j$ we have one
plane wave, in condition (2.5) we have two (one -- $k=0$ -- fast,
other -- $k=1$ --- slow), and in condition (2.6) the direct plane
waves are absent. The polarization vectors ${\bf p}_j$ of plane
waves in the problems $A$ and $B$ are identical, with the
exception of the case (2.6), besides in the case (2.5) both fast
and slow waves have the same polarization vector. Henceforth we
denote the phase velocity in the cases (2.4), (2.5) by uniform
way: $\cb = \cbn = (-1)^k \can - w_n$, $(k=0)\vee(k=0,1)$, i.e.
for the case (2.4) $c^B_{pj} = c^{B(0)}_{pj}$.

If for $\forall {\bf n}$, $\forall j \in \{1,2\}$ the condition
(2.4) is realized, then we shall call subseismic the motion
behavior, and otherwise -- trans- or superseismic. The difference
between that notions we shall mark below.

We introduce the phase velocity vector $\vcb$ and the inverse
velocity vector $\vLb$
$$
\vcb = \cb{\bf n} = (-1)^k \vcan - w_n{\bf n}; \quad \vcan =
\can{\bf n}, \eqno(2.7)
$$
$$
\vLb={\bf n}/\cb. \eqno(2.8)
$$

By analogy with problem $A$, we shall also call the vector $\vLb$
refraction vector or slowness vector.

The dispersion equation in the problem $B$ for plane wave (2.1)
have the following form
$$
D_B (\balpha , \omega ) = {\rm det} \, [{\bf \Gamma} (\balpha ) -
\rho \Omega^2 (\balpha) {\bf E}] = 0 , \eqno(2.9)
$$
where ${\bf E}$ is the unit matrix 2x2,
$$
\Omega (\balpha)=\omega +{\bf w} \cdot \balpha. \eqno(2.10)
$$

The solution of equation (2.10) can be leaved in the form of the
set of hypersurfaces
$$
\omega= \omega^{B(k)}_j (\balpha ) = \alpha \cbn , \quad j=1,2.
 \eqno(2.11)
$$

From the dispersion surface (2.11) the group velocity vector
$\vcgb = \vcgb ({\bf n})$ may be found by the following way
$$
\vcgb = \frac{\partial \omega^{B(k)}_j(\balpha)}{\partial \balpha}
= \vcb + ({\bf E} - {\bf n}{\bf n}^*) \cdot \frac{\partial
\cb}{\partial {\bf n}}, \eqno(2.12)
$$
or, by using (2.7),
$$
\vcgb = (-1)^k \vcgan - {\bf w}. \eqno(2.13)
$$

It is known [10], that for plane waves in the problem $A$ these
important relations are correct \quad (${\bf L}^A={\bf n}/c^A_p$)
$$
{\bf c}^A_g \cdot {\bf n} = c^A_p, \quad {\bf c}^A_g \cdot {\bf
L}^A = 1, \quad {\bf n} \cdot d{\bf c}^A_g =0, \quad {\bf c}^A_g
\cdot d{\bf L}^A = 0.
$$

As it is shown in [7] for the case of tree-dimensional problem,
the analogical formulae are valid. This proof is completely
transported on the considered plane problem. Therefore the
following relations are correct
$$
\vcgb \cdot {\bf n} = c^{B(k)}_{pj}; \; \vcgb \cdot \vLb = 1; \;
{\bf n} \cdot d \vcgb =0; \; \vcgb \cdot d \vLb = 0. \eqno(2.14)
$$
(In (2.14) and in the next the summation by repeating index is
absent.)

After analyzing obtained formulae (2.14), we may establish for the
problem $B$ the basic properties of phase velocity curves $\vcb
({\bf n})= \cbn {\bf n}$, slowness curves (inverse velocity or
reflection curves) $\vLb ({\bf n})= {\bf n} /\cbn$ and group
velocity curves (waves curves or ray curves) $\vcgb ({\bf n})$. In
general, all of these curves in the problem $B$ with $w \neq 0$ do
not have the central symmetry ${\bf n} \leftrightarrow (-{\bf
n})$, and their crystallographical symmetry does not reserve. The
phase velocity curves $\vcb$ and waves curves $\vcgb$ are limited
for all values $w$. But the slowness curves $\vLb$ can be both
limited and unlimited depending on motion behavior.

We shall call the source motion rate superseismic, if in ${\rm
R}^2$ the directions  ${\bf n}$, along which there are no slowness
curves, exist, and all slowness curves are unlimited. If there is
one limited slowness curve, but also unlimited slowness curves
exist, then we shall call this motion behavior transseismic.

We note that the slowness curves $\vLb$ may essentially differ
from corresponding curves for the problem $A$, and their number
may vary from 2 to 4. At the same time, the group velocity curves
$\vcgb$ for the problem $B$ may be obtained by simple transfer of
curves for the problem $A$ by vector $(-{\bf w})$. Besides, with
trans- and superseismic motion rate two parts ${\bf c}^{B0}_{gj}$
and ${\bf c}^{B1}_{gj}$ form one closed curve. For this behavior
in case (2.5) for fixed ${\bf n}$ and $j$ we have two plane waves
(fast and slow) with inverse velocities, which belong to two
different slowness curves.

In addition to this from (2.2) and (2.14) important common
properties follow [7]. Thus, the quasi-longitudinal and
quasi-shear curves save their types for all ${\bf w}$, because the
polarization vector  ${\bf p}_j$ from (2.2) does not depend on the
velocity ${\bf w}$.

The group velocity vector ${\bf c}_g$, by (2.14) in every point of
slowness curve ${\bf L}$, is orthogonal to the tangent to ${\bf
L}$ (${\bf c}_g \cdot d{\bf L} =0$). Conversely, the wave normal
vector ${\bf n}$ of plane wave with group velocity ${\bf c}_g
({\bf n})$ is orthogonal to tangent in the corresponding point of
wave velocity ${\bf c}_g$ (${\bf n} \cdot d{\bf c}_g =0$).

% ******************New text**********************

Everything considered above is illustrated by Figs.~1--11. These
figures show slowness and group velocities curves for different
kinds of quartz for plane deformations in planes $Ox_1x_2$ and
$Ox_2x_3$. Curves marked with "1" and "2" correspond to waves with
phase velocities $c^{B(k)}_{p1}$ and $c^{B(k)}_{p2}$ respectively.
The waves are numbered so that for the problem $A$
$c^A_{p1}<c^A_{p2}$. Therefore in the most cases waves with
subscript "1" will be quasi-shear and waves with subscript "2"
will be quasi-longitudinal.

The figures 1--6 show characteristic curves of plane waves for
fused silica ($SiO_2$), which is isotropic material. The
elasticity modules and density of fused silica as well as other
materials were taken from [9].

The figures 1 and 2 show curves of slowness and group velocities
of fused silica without motion ($w=0$) respectively. As it can be
seen from Fig.~1 and 2 when $w=0$ these curves represent couples
of concentric circumferences.

When the velocity is not equal to zero pictures of characteristic
curves essentially change. For subseismic motion when
$w_1=0.6\cdot 10^3$ m/s; $w_2=0$ the curves slowness and group
velocities are shown on Fig.~3, 4 respectively. Comparing Fig.~1
and Fig.~3 we can notice that the curves of slowness change their
shape and structure when $w$ increases. Meantime, as it was
mentioned above, group velocity curves (Fig.~2 and Fig.~4) simply
transfer on vector $-{\bf w}$ relatively to the center of
coordinate system.

\includegraphics[bb = -35pt 0pt 780pt 340pt, scale=0.5]{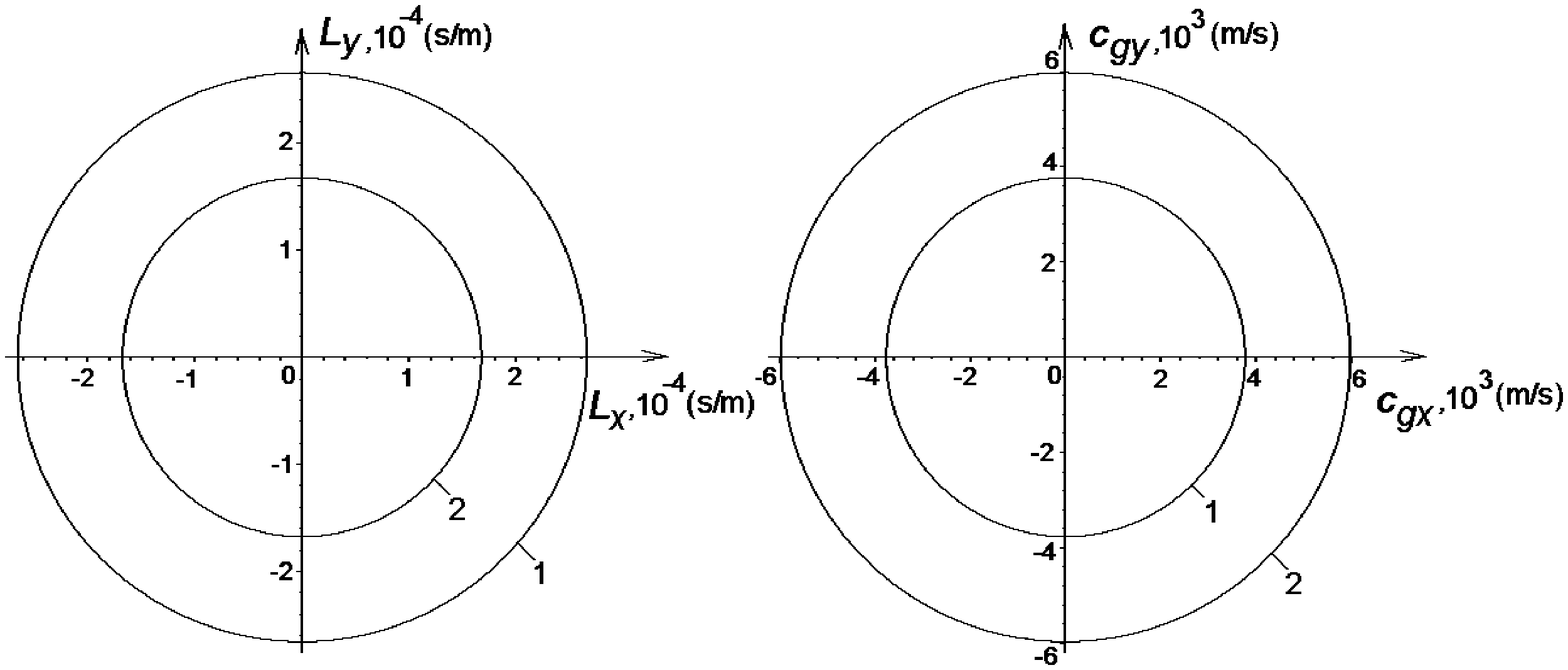}
\begin{center}
{Fig.~1. Slowness curves, \qquad \qquad Fig.~2. Group
velocity curves, \qquad\\
fused silica, $w_1=w_2=0$. \qquad \qquad \qquad fused silica,
$w_1=w_2=0$.}
\end{center}

\includegraphics[bb = -35pt 10pt 780pt 320pt, scale=0.5]{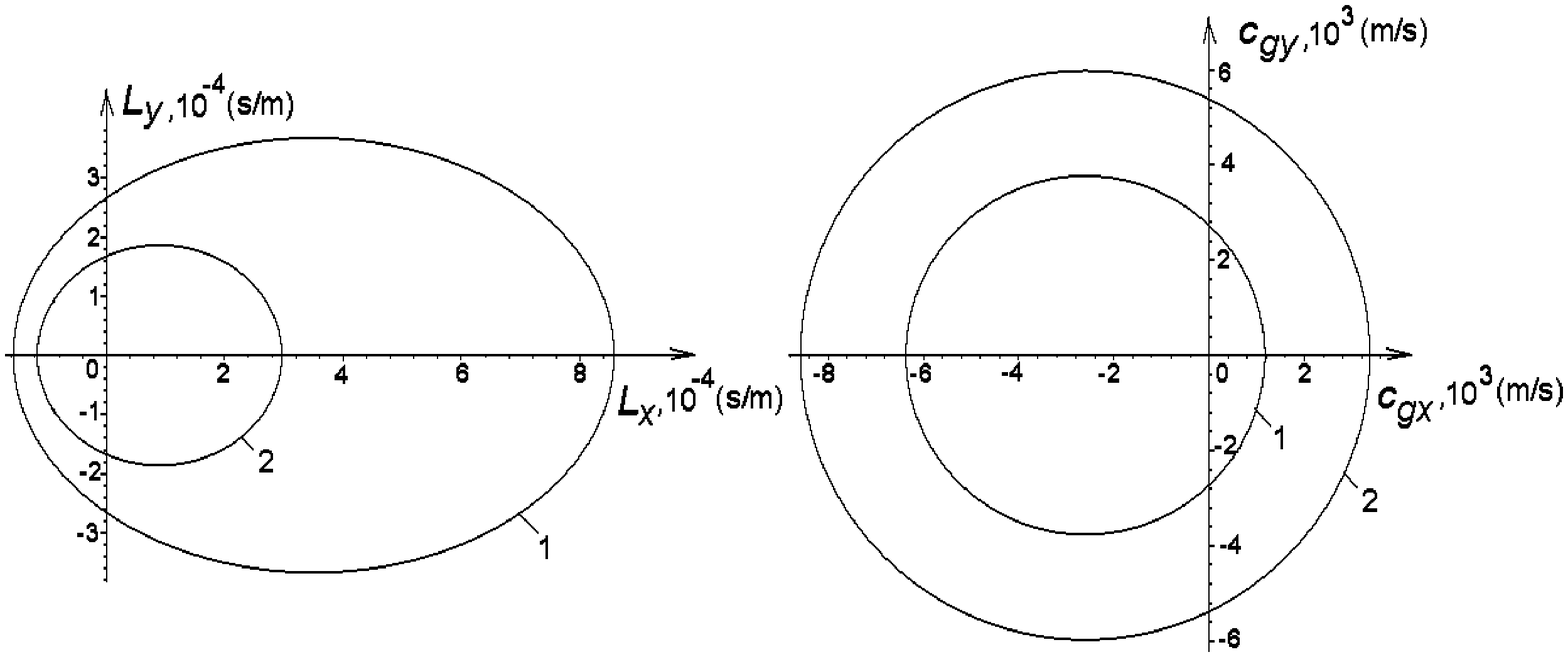}
\begin{center}
{Fig.~3. Slowness curves, \qquad \qquad Fig.~4. Group
velocity curves, \qquad\\
\quad fused silica, $w_1=2600$, $w_2=0$. \qquad fused silica,
$w_1=2600$, $w_2=0$.}
\end{center}

\includegraphics[bb = -35pt 0pt 780pt 320pt, scale=0.5]{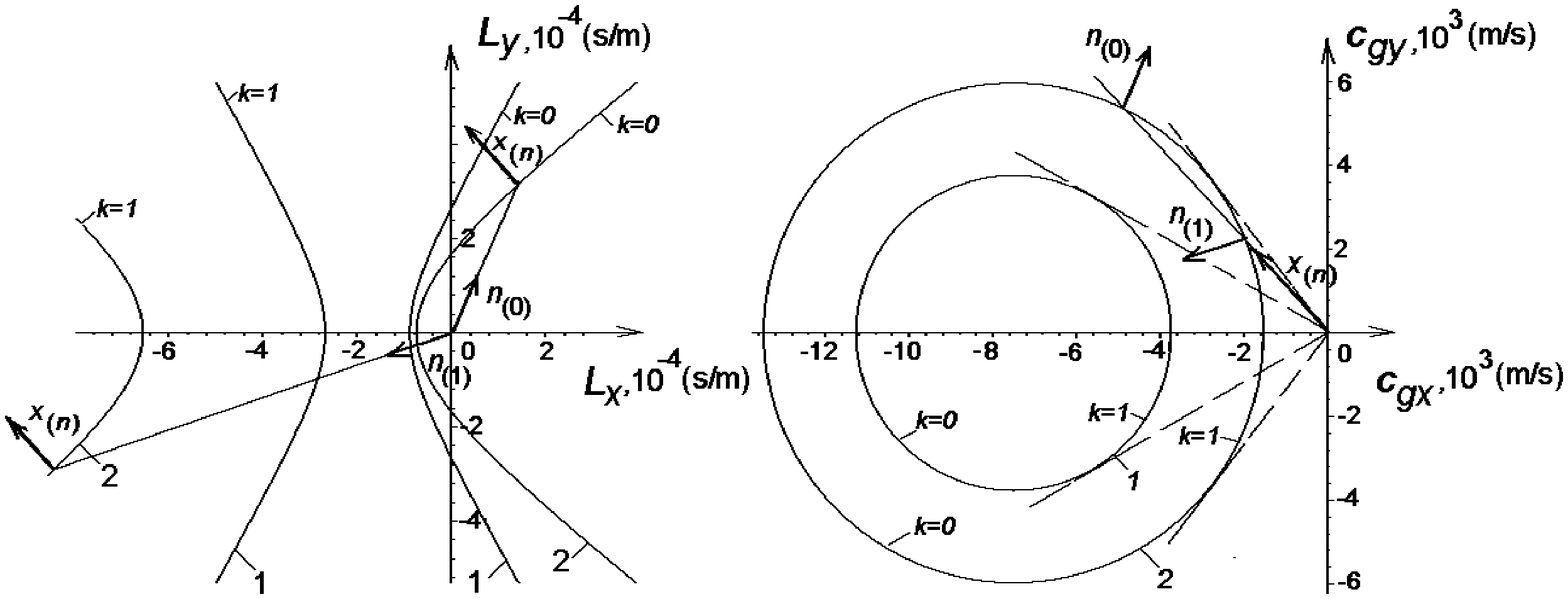}
\begin{center}
{Fig.~5. Slowness curves, \qquad \qquad Fig.~6. Group
velocity curves, \qquad\\
\quad fused silica, $w_1=7500$, $w_2=0$. \qquad fused silica,
$w_1=7500$, $w_2=0$.}
\end{center}

The changes of the shape and structure of slowness curves and
ultimate tendencies of these changes can be seen from Fig.~5, 6,
where characteristic curves for superseismic motion when
$w_1=7.5\cdot 10^3$ m/s; $w_2=0$ are shown. Here Fig.~5 shows
slowness curves and Fig.~6 shows group velocity curves. Slowness
curves for superseismic motion (Fig.~5) are not limited. Each
couple of these curves corresponds to fast and slow waves. Group
velocity curves (Fig.~6) when $w_1=7.\cdot 10^3$ m/s; $w_2=0$
wholly move to the half-plane $x \le 0$, this is caused by the
absence of waves in front of the source for superseismic motion.
The parts of fast and slow waves for group velocity curves are the
pieces of the same curves separated by tangents to these curves
which pass through the center of coordinate system (Fig.~6).

The figures 5 and 6 illustrate relations of orthogonality ${\bf
c}_g \cdot d{\bf L} =0$ and ${\bf L} \cdot d {\bf c}_g ={\bf n}
\cdot d {\bf c}_g =0$. Here slowness curves have directions ${\bf
n}_{(0)}$ and ${\bf n}_{(0)}$ associated with one direction ${\bf
x}_{(n)}$ on picture of group velocities, and  vice versa.

For anisotropic materials the pictures of characteristic curves
can be much more complicate. The Figures ~7--10 show curves for
slowness and group velocities curves for $\alpha$-quartz ($SiO_2$)
for plane deformation in the plane $x_2x_3$. This material relates
to trigonal crystal system of 32 class and possesses piezoelectric
properties which we don't take into consideration.

\includegraphics[bb = -35pt 0pt 780pt 320pt, scale=0.5]{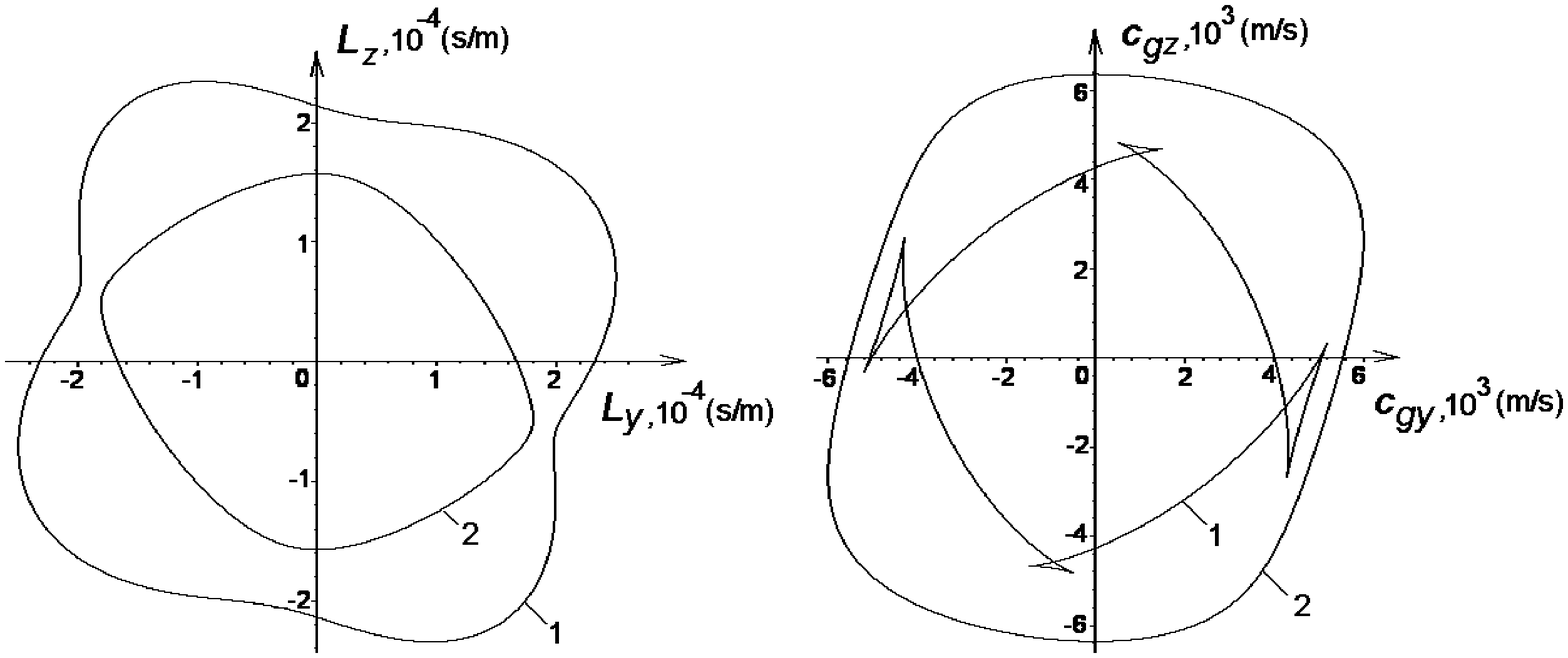}
\begin{center}
{Fig.~7. Slowness curves, \qquad \qquad Fig.~8. Group
velocity curves, \qquad\\
\quad $\alpha$-quartz, $w_1=w_2=0$. \quad \qquad $\alpha$-quartz,
$w_1=w_2=0$.}
\end{center}

\includegraphics[bb = -35pt 0pt 780pt 320pt, scale=0.5]{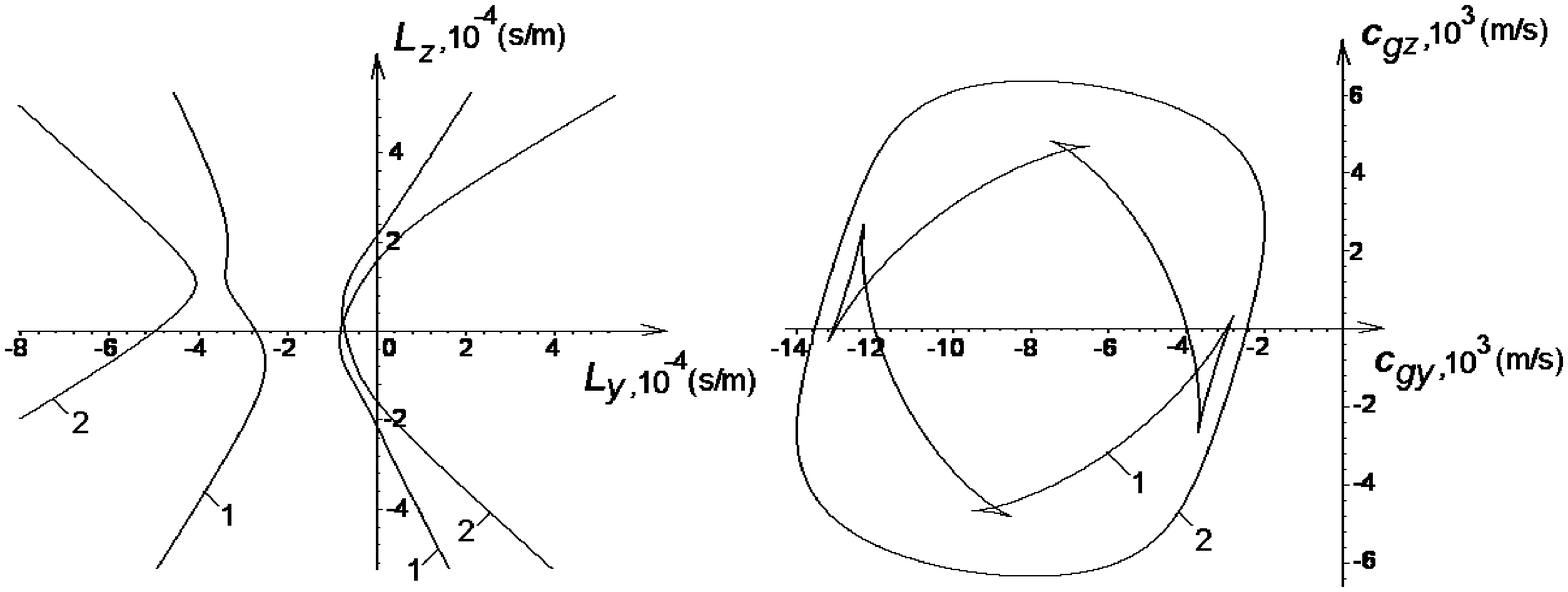}
\begin{center}
{Fig.~9. Slowness curves, \qquad \qquad Fig.~10. Group velocity
curves, \qquad\\
\quad $\alpha$-quartz, $w_1=8000$, $w_2=0$. \qquad
$\alpha$-quartz, $w_1=8000$, $w_2=0$.}
\end{center}

Slowness curves for $\alpha$-quartz on Fig.~7 and 9 have the
points of inflection and the sections of convexity and concavity.
Therefore group velocity curves have typical acute edges.
Availability of such edges defines areas with different numbers of
spreading waves even for problem $A$ [11]. This property is also
secured for problem $B$ with additional sophistication of the wave
picture and distraction of the curve central symmetry.

% ******************End of new text**********************

\section{Fundamental solutions}

The fundamental solution for the problem $B$ is the solution of
equation (1.9) with the point external source ${\bf f}$
$$
{\bf f}=f \delta ({\bf x}) {\bf l} \, \exp{(i\omega t)}.
$$

For obtaining unique solution we use the principle of limiting
absorption, according to which in the case of oscillation of point
source by the law $\exp{(i\omega t)}$ the amplitude ${\bf v} ({\bf
x})$ in equation (1.1) should be defined as a limit at
$\varepsilon \to +0$
$$
{\bf v} = \lim_{\varepsilon \to +0} {\bf v}_\varepsilon.
\eqno(3.1)
$$
The function ${\bf v}_\varepsilon$ satisfies the following
$\varepsilon$-problem
$$
-{\bf L}^* (\nabla^x) \cdot {\bf C} \cdot {\bf L}(\nabla^x) \cdot
{\bf v}_{\varepsilon} + \rho (i\omega_{\varepsilon} - {\bf w}
\cdot \nabla^x)^2 {\bf v}_{\varepsilon} = f \delta ({\bf x}) {\bf
l}, \eqno(3.2)
$$
where $\omega_{\varepsilon}=\omega -i\varepsilon$, $\varepsilon
> 0$, $\varepsilon \ll 1$.

By applying Fourier integral transforms along $x_1$, $x_2$, we
obtain the fundamental solution ${\bf v}_\varepsilon$ for equation
(3.2) in integral form
$$
{\bf v}_\varepsilon ({\bf x}) = \frac{f}{4\pi^2} \int \!\!\!
\int_{R^2} {\bf K}^{-1}(\balpha)\cdot {\bf l} \, {\rm
e}^{-i\balpha \cdot {\bf x}} \, d\alpha_1 \, d\alpha_2, \eqno(3.3)
$$
where
$$
{\bf K}(\balpha) = {\bf \Gamma}(\balpha) - \rho
\Omega^2_\varepsilon (\balpha) {\bf E}, \eqno(3.4)
$$
$$
\Omega_\varepsilon (\balpha) = \omega_\varepsilon + {\bf w} \cdot
\balpha.  \eqno(3.5)
$$

Introducing the polar coordinate system ($\alpha$, $\theta$)
$$
\balpha = \alpha {\bf n}, \quad {\bf n} = \{ \cos\theta ; \,
\sin\theta \}, \eqno(3.6)
$$
it is possible to decompose the matrix ${\bf K}^{-1} (\balpha)$ by
accompanying matrices ${\bf P}_j$ [12], which are composed of
eigenvectors ${\bf p}_1$, ${\bf p}_2$ for problem (2.2)
$$
{\bf K}^{-1}(\balpha) = \frac1\rho \sum_{j=1}^2\frac{{\bf P}_j}
{\alpha^2\nu^2_j - \Omega^2_\varepsilon (\balpha)}, \eqno(3.7)
$$
$$
{\bf P}_j ={\bf P}_j({\bf n}) = {\bf p}_j({\bf n}){\bf p}^*_j({\bf
n}), \quad {\bf n}={\bf n} (\theta ), \eqno(3.8)
$$
with regard to (3.6)---(3.8) we obtain from (3.3)
$$
{\bf v}_\varepsilon  = \frac{f}{4\pi^2\rho} \sum_{j=1}^2
\int_0^{2\pi} {\bf P}_j \cdot {\bf l} \, I_{j\varepsilon} \,
d\theta, \eqno(3.9)
$$
$$
I_{j\varepsilon} = \int_0^{+\infty} \frac{\alpha} {\alpha^2\nu^2_j
({\bf n}) - \Omega^2_\varepsilon (\alpha {\bf n})} \, {\rm
e}^{-iz\alpha} \, d\alpha, \quad z={\bf n} \cdot {\bf x}.
\eqno(3.10)
$$

We decompose integrand function from (3.10) by partial fractions
$$
\frac{\alpha}{\alpha^2\nu^2_j ({\bf n}) - \Omega^2_\varepsilon
(\alpha {\bf n})} = \frac{1}{2\omega_\varepsilon \nu_j} \left(
\frac{\alpha^+_{j\varepsilon}} {\alpha - \alpha^+_{j\varepsilon}}-
\frac{\alpha^-_{j\varepsilon}} {\alpha - \alpha^-_{j\varepsilon}}
\right), \eqno(3.11)
$$
$$
\alpha^+_{j\varepsilon} = \frac{\omega_{\varepsilon}}{\nu_j-w_n} ,
\quad \alpha^-_{j\varepsilon} = -
\frac{\omega_{\varepsilon}}{\nu_j+w_n}. \eqno(3.12)
$$

In the integral (3.9) from $\theta$ we realize the following
transformations
$$
\int\limits^{2\pi}_0 (...)(\theta)\, d\theta = \int\limits^{\pi/2+
\tilde \theta}_{-\pi/2+ \tilde \theta} (...)(\theta)\, d\theta +
\int\limits^{\pi/2+ \tilde \theta}_{-\pi/2+ \tilde \theta}
(...)(\theta +\pi)\, d\theta, \eqno(3.13)
$$
where $\tilde{\theta}$ is the angle in the polar coordinate system
for physical plane ${\bf x}$
$$
{\bf x} =r{\bf y}, \quad |{\bf y}|=1, \quad {\bf y}=\{ \cos \tilde
\theta, \sin \tilde \theta \}. \eqno(3.14)
$$

By taking into account (3.11)---(3.13) the formulae (3.9), (3.10)
can be transformed in the following way
$$
{\bf v}_\varepsilon ({\bf x}) = \frac{f}{8\pi^2\rho
\omega_\varepsilon} \sum^2_{j=1} \int\limits^{\pi/2+ \tilde
\theta}_{-\pi/2+ \tilde \theta} \frac{{\bf P}_j ({\bf n}) \cdot
{\bf l}}{\nu_j ({\bf n})} \, J_{j\varepsilon} (\theta) \, d\theta,
\eqno(3.15)
$$
$$
J_{j\varepsilon} (\theta) = \alpha^+_{j\varepsilon} \left(
\int^{+\infty}_0 \frac{{\rm e}^{-iz\alpha}} {\alpha -
\alpha^+_{j\varepsilon}} \, d\alpha + \int^{+\infty}_0 \frac{{\rm
e}^{iz\alpha}} {\alpha + \alpha^+_{j\varepsilon}} \, d\alpha
\right) - \eqno(3.16)
$$
$$
-\alpha^-_{j\varepsilon} \left( \int^{+\infty}_0 \frac{{\rm
e}^{-iz\alpha}} {\alpha - \alpha^-_{j\varepsilon}} \, d\alpha +
\int^{+\infty}_0 \frac{{\rm e}^{iz\alpha}} {\alpha +
\alpha^-_{j\varepsilon}} \, d\alpha \right).
$$

The calculation of integrals from (3.16) depends on real and
imaginary parts of $\alpha^{\pm}_{j\varepsilon}$, which in their
turn depend on moving rate
$$
{\rm Re} \, \alpha^+_{j\varepsilon} >0, \; {\rm Im} \,
\alpha^+_{j\varepsilon} \leq 0, \quad \nu_j-w_n>0,
$$
$$
{\rm Re} \, \alpha^+_{j\varepsilon} <0, \; {\rm Im} \,
\alpha^+_{j\varepsilon} \geq 0, \quad \nu_j-w_n<0, \eqno(3.17)
$$
$$
{\rm Re} \, \alpha^-_{j\varepsilon} <0, \; {\rm Im} \,
\alpha^-_{j\varepsilon} \geq 0, \quad \nu_j+w_n>0,
$$
$$
{\rm Re} \, \alpha^-_{j\varepsilon} >0, \; {\rm Im} \,
\alpha^-_{j\varepsilon} \leq 0, \quad \nu_j+w_n<0.
$$

Besides, it is essentially, that $\forall \theta \in
[-\pi/2+\tilde{\theta}, \pi/2+\tilde{\theta}]$
$$
z={\bf n} \cdot {\bf x} = r \cos (\theta-\tilde{\theta}) > 0.
\eqno(3.18)
$$

We shall introduce the contours $C^{\pm}_{\Gamma}=[0,R] \cup
C^{\pm}_R \cup [\pm iR,0]$, where $C^+_R$ and $C^-_R$ are the
quoters of circle with radius $R$ and center in the origin of
coordinate system, lying respectively in the first and in the
fourth quadrant of complex plane $\alpha$, $R\gg 1$.

Calculating the integrals by complex integration method, using
contours $C^{\pm}_{\Gamma}$, relations (3.17), (3.18) and Gordan's
lemma and turned $R$ to infinity, we obtain ($z>0$)
$$
\int^{+\infty}_0 \frac{{\rm e}^{-iz\alpha}} {\alpha -
\alpha^{\pm}_{j\varepsilon}} \, d\alpha = -2\pi i {\rm e}^{-iz
\alpha^{\pm}_{j\varepsilon}} H[\pm \nu_j -w_n ] +i\int^{+\infty}_0
\frac{(\alpha^{\pm}_{j\varepsilon}-i\eta)}
{(\alpha^{\pm}_{j\varepsilon})^2+\eta^2} {\rm e}^{-z\eta} \,
d\eta, \eqno(3.19)
$$
$$
\int^{+\infty}_0 \frac{{\rm e}^{iz\alpha}} {\alpha +
\alpha^{\pm}_{j\varepsilon}} \, d\alpha = i\int^{+\infty}_0
\frac{(\alpha^{\pm}_{j\varepsilon}-i\eta)}
{(\alpha^{\pm}_{j\varepsilon})^2+\eta^2} {\rm e}^{-z\eta} \,
d\eta,
$$
where $H$ is the Heaviside function.

On substitution (3.19) into (3.16) and received formula into
(3.15) and implementing passage to the limit when $\varepsilon \to
+0$, we obtain the final representation of fundamental solutions
for the problems $A$ and $B$ with arbitrary moving rate
$$
{\bf v}={\bf v}_d+{\bf v}_0, \eqno(3.20)
$$
$$
{\bf v}_d= \frac{if}{4\pi \rho} \sum^2_{j=1} ({\sum_k}')
(-1)^{k+1}I_{jk}, \eqno(3.21)
$$
$$
I_{jk}=\int\limits^{\pi/2+ \tilde \theta}_{-\pi/2+ \tilde \theta}
\frac{{\bf P}_j (\theta)\cdot {\bf l} \; H[(-1)^k\nu_j-w_n]}{\nu_j
(\theta) c^{B(k)}_{pj}(\theta )} \, {\rm e}^{-i\omega {\bf
L}^{B(k)}_j \cdot {\bf x}} \, d\theta, \eqno(3.22)
$$
$$
{\bf v}_0=\frac{if}{4\pi^2 \rho \omega} \sum^2_{j=1}
\int\limits^{\pi/2+ \tilde \theta}_{-\pi/2+ \tilde \theta}
\frac{{\bf P}_j (\theta)\cdot {\bf l}} {\nu_j (\theta)} \, I^0_j
\, d\theta, \eqno(3.23)
$$
$$
I^0_j=\int\limits^{+\infty}_0 \left( \alpha^+_j \frac{\alpha^+_j
-i\eta}{(\alpha^+_j)^2+\eta^2} - \alpha^-_j \frac{\alpha^-_j
-i\eta}{(\alpha^-_j)^2+\eta^2} \right) \, {\rm e}^{-z\eta} \,
d\eta, \eqno(3.24)
$$
$$
z={\bf n}\cdot {\bf x}=r \cos (\theta - \tilde \theta), \quad
\alpha^{\pm}_j = \omega/(\pm \nu_j-w_n), \eqno(3.25)
$$
where symbol $\sum_k^{\prime}$ in (3.21) denotes the presence or
the absence of summation by $k$ according to moving rate.

In addition formula (3.12) can be transformed by using the
following integrals
$$
\int^{+\infty}_0 \frac{{\rm e}^{-z\eta }}{\zeta^2+\eta^2}\, \eta
\, d\eta = -[{\rm ci}\, (\zeta z) \cos (\zeta z)+ {\rm si}\,
(\zeta z) \sin (\zeta z)], \eqno(3.26)
$$
$$
\int^{+\infty}_0 \frac{{\rm e}^{-z\eta }}{\zeta^2+\eta^2}\,  d\eta
= \frac1{\zeta} [{\rm ci}\, (\zeta z) \sin (\zeta z)- {\rm si}\,
(\zeta z) \cos (\zeta z)] ,
$$
where ${\rm si}\, (x)$ and ${\rm ci}\, (x)$ are respectively
integral sine and cosine.

For the problem $A$ with $w=0$ the fundamental solutions
representations in form (3.21)---(3.26) can be significantly
simplified
$$
{\bf v}_d = \frac{if}{4\pi \rho} \sum^2_{j=1} \int\limits^{\pi/2+
\tilde \theta}_{-\pi/2+ \tilde \theta} \frac{{\bf P}_j
(\theta)\cdot {\bf l}} {\nu^2_j (\theta)} \, {\rm e}^{-i\omega
{\bf L}^A_j \cdot {\bf x}} \, d\theta, \eqno(3.27)
$$
$$
{\bf v}_0=-\frac{if}{2\pi^2 \rho} \sum^2_{j=1} \int\limits^{\pi/2+
\tilde \theta}_{-\pi/2+ \tilde \theta} \frac{{\bf P}_j
(\theta)\cdot {\bf l}} {\nu^2_j (\theta)} \, [{\rm ci}\,
(\frac{\omega}{\nu_j} z) \cos (\frac{\omega}{\nu_j} z)+ {\rm si}\,
(\frac{\omega}{\nu_j} z) \sin (\frac{\omega}{\nu_j} z)] \,
d\theta.
$$

Thus, we have obtained the fundamental solutions for problem $A$
and $B$ in the form of integrals by finite interval. It should be
noted that the representation (3.20)---(3.26) for the problem $B$
are suitable for arbitrary moving rate.

\section{Far field asymptotics}

The term ${\bf v}_d$ from (3.21), (3.22) or (3.27) define dynamic
effect of displacement field. We shall determine the asymptotics
of ${\bf v}_d$ in far field $\omega r >> 1$ for general case of
the problem $B$. For this it is evidently required to find the
asymptotics of oscillating integral (3.22).

From classical method of stationary phase the contribution of
individual stationary point $\theta_s$ for integral
$$
\int F(\theta ) {\rm e}^{iq(\theta )\omega r} \, d\theta
\eqno(4.1)
$$
at $\omega r >>1$ can be given by the expression
$$
\sqrt{\frac{2\pi}{\omega r|q''(\theta_s)|}}
 F(\theta_s) {\rm e}^{i[q(\theta_s)\omega r+
\frac{\pi}4 {\rm sign}\, q''(\theta_s)]} \, d\theta, \eqno(4.2)
$$
where $\theta_s$ is the saddle (stationary) point, or the root of
equation
$$
q'(\theta_s)=0. \eqno(4.3)
$$

In our problem for fixed $j$ and $k$ we have
$$
F(\theta )=\frac{{\bf P}_j (\theta)\cdot {\bf l} \;
H[(-1)^k\nu_j-w_n]} {\nu_j (\theta) c^{B(k)}_{pj} (\theta)} ,
\quad q(\theta)=- {\bf L}^{B(k)}_j \cdot {\bf y}. \eqno(4.4)
$$

We shall obtain the formula for saddle point $\theta_s$. From
(4.3), (4.4) we have
$$
\frac{\partial q}{\partial \theta} = - \frac{\partial {\bf
L}^{B(k)}_j}{\partial \theta} \cdot {\bf y}. \eqno(4.5)
$$

We can represent the components of vector $\partial {\bf
L}^{B(k)}_j / \partial \theta$ in the form
$$
\frac{\partial L^{B(k)}_{ji}}{\partial \theta} = \sum^2_{m=1}
\frac{\partial L^{B(k)}_{ji}}{\partial n_m} \; \frac{\partial
n_m}{\partial \theta},
$$
and since
$$
\frac{\partial L^{B(k)}_{ji}}{\partial n_m} =
\frac{\partial}{\partial n_m}(\frac{n_i}{c^{B(k)}_{pj}}) =
\frac1{c^{B(k)}_{pj}} (\delta_{im} - L^{B(k)}_{ji} \frac{\partial
c^{B(k)}_{pj}}{\partial n_m}),
$$
it follows
$$
\frac{\partial {\bf L}^{B(k)}_j}{\partial \theta} =
\frac1{c^{B(k)}_{pj}} [{\bf E} - {\bf L}^{B(k)}_j (\frac{\partial
c^{B(k)}_{pj}}{\partial {\bf n}})^*] \cdot \frac{\partial {\bf
n}}{\partial \theta}. \eqno(4.6)
$$

Then by using (2.12) we find
$$
{\bf c}^{B(k)}_{gj} \cdot \frac{\partial {\bf n}}{\partial \theta}
= \frac{\partial c^{B(k)}_{pj}}{\partial {\bf n}} \cdot
\frac{\partial {\bf n}}{\partial \theta}, \eqno(4.7)
$$
since ${\bf c}^{B(k)}_{pj} \cdot (\partial {\bf n}/\partial \theta
) = 0$, ${\bf n} \cdot (\partial {\bf n}/\partial \theta ) =0$.

From (4.7) the formula (4.6) can be rewritten in the form
$$
\frac{\partial {\bf L}^{B(k)}_j}{\partial \theta} =
\frac1{c^{B(k)}_{pj}} [{\bf E} - {\bf L}^{B(k)}_j ({\bf
c}^{B(k)}_{gj})^*] \cdot \frac{\partial {\bf n}}{\partial \theta}.
\eqno(4.8)
$$

On substitution of (4.8) into (4.5), (4.3), we obtain
$$
{\bf y}^* \cdot [{\bf E} - {\bf L}^{B(k)}_j ({\bf
c}^{B(k)}_{gj})^*] \cdot \frac{\partial {\bf n}}{\partial \theta}
=0. \eqno(4.9)
$$
Besides, using one of the relations (2.14) $({\bf L}^{B(k)}_j
\cdot {\bf c}^{B(k)}_{gj} =1)$ we can write following sequence of
equalities
$$
0= {\bf n} \cdot {\bf y} - ({\bf L}^{B(k)}_j \cdot {\bf
c}^{B(k)}_{gj}) {\bf n} \cdot {\bf y} = {\bf n}^* \cdot {\bf E}
\cdot {\bf y} - ({\bf n} \cdot {\bf c}^{B(k)}_{gj}) \frac{{\bf
n}}{c^{B(k)}_{pj}} \cdot {\bf y} =
$$
$$
= {\bf n}^* \cdot {\bf E} \cdot {\bf y} - {\bf n}^* \cdot ({\bf
c}^{B(k)}_{gj} ({\bf L}^{B(k)}_j)^*) \cdot {\bf y} = {\bf y}^*
\cdot [{\bf E} - {\bf L}^{B(k)}_j ({\bf c}^{B(k)}_{gj})^* ] \cdot
{\bf n},
$$
i.e.
$$
{\bf y}^* \cdot [{\bf E} - {\bf L}^{B(k)}_j ({\bf
c}^{B(k)}_{gj})^* ] \cdot {\bf n} = 0. \eqno(4.10)
$$

From (4.9), (4.10) we obtain, that the vector ${\bf y}^* \cdot
[{\bf E} - {\bf L}^{B(k)}_j ({\bf c}^{B(k)}_{gj})^* ]$ is
orthogonal for two one-to-one orthogonal vectors ${\bf n}$ and
$\partial {\bf n}/\partial \theta$, and therefore, it is equal to
zero
$$
{\bf y}^* \cdot [{\bf E} - {\bf L}^{B(k)}_j ({\bf
c}^{B(k)}_{gj})^* ] = 0. \eqno(4.11)
$$

By using inequality (3.18), which can be rewritten in the form
$$
{\bf y} \cdot {\bf L}^{B(k)}_j (\theta)>0, \eqno(4.12)
$$
we can get the following shape for the equation (4.11)
$$
{\bf c}^{B(k)}_{gj}(\theta) \, |{\bf y} \cdot {\bf L}^{B(k)}_j
(\theta)|= {\bf y}. \eqno(4.13)
$$

The relations (4.13), (4.12) are the suitable formulae for
definitions of stationary points $\theta_{jm}^{(k)}$ and wave
normal vector ${\bf n}_{jm}^{(k)}={\bf n}(\theta_{jm}^{(k)})$.

Geometrically, from (4.12), (4.13), just as in problem $A$, for
stationary points the group velocity vector ${\bf
c}^{B(k)}_{gjm}={\bf c}^{B(k)}_{gj} ({\bf n}_{jm}^{(k)})$, which
is perpendicular to slowness curve ${\bf L}^{B(k)}_j$, is directed
along vector ${\bf y}$ (${\bf x}$). Therefore, for fixed direction
${\bf x}$ on physical plane the stationary value ${\bf
n}_{jm}^{(k)}$ will appear such values ${\bf n}$, in which the
external normal to slowness curve coincides with direction ${\bf
x}$.

Now, we shall dwell upon calculation the expression $q''$ in the
stationary point $\theta_{jm}^{(k)}$
$$
q''=-{\bf y}\cdot \frac{\partial^2 {\bf L}^{B(k)}_j}{\partial
\theta^2}. \eqno(4.14)
$$

Let $\btau$ be tangent to slowness curve ${\bf L}^{B(k)}_j$ unit
vector and $s$ be natural parameter along this curve. We note,
that ${\bf y}$ is the unit normal vector to ${\bf L}^{B(k)}_j$,
and thus, ${\bf y}\cdot \partial {\bf L}^{B(k)}_j/\partial \theta
=0$.

Because
$$
\frac{\partial {\bf y}}{\partial \theta} \cdot \frac{\partial {\bf
L}^{B(k)}_j}{\partial \theta} = \frac{\partial}{\partial \theta}
\left( {\bf y}\cdot \frac{\partial {\bf L}^{B(k)}_j}{\partial
\theta} \right)- {\bf y}\cdot \frac{\partial^2 {\bf
L}^{B(k)}_j}{\partial \theta^2},
$$
then we can transform formulae (4.14) in the form
$$
q''=\frac{\partial {\bf y}}{\partial \theta}\cdot \frac{\partial
{\bf L}^{B(k)}_j}{\partial \theta}. \eqno(4.15)
$$

By using Frene formulae, we can write
$$
\frac{\partial {\bf L}^{B(k)}_j}{\partial \theta}= \frac{\partial
{\bf L}^{B(k)}_j}{\partial s} \frac{\partial s}{\partial \theta}=
\btau \frac{\partial s}{\partial \theta} , \quad \frac{\partial
{\bf y}}{\partial \theta}= \frac{\partial {\bf y}}{\partial s}
\frac{\partial s}{\partial \theta}= -k^{B(k)}_{pj}\btau
\frac{\partial s}{\partial \theta}, \eqno(4.16)
$$
where $k^{B(k)}_{pj}$ is the curvature of slowness curve ${\bf
L}^{B(k)}_j$.

From (4.15), (4.16) we have
$$
q''=-k^{B(k)}_{pj} \left( \frac{\partial s}{\partial \theta}
\right)^2= -k^{B(k)}_{pj}\frac{\partial {\bf L}^{B(k)}_j}{\partial
\theta} \cdot \frac{\partial {\bf L}^{B(k)}_j}{\partial \theta}.
\eqno(4.17)
$$

By using (4.8), it is easy to find, that
$$
\frac{\partial {\bf L}^{B(k)}_j}{\partial \theta} \cdot
\frac{\partial {\bf L}^{B(k)}_j}{\partial \theta} =
\frac1{(c^{B(k)}_{pj})^4} [(c^{B(k)}_{pj})^2+({\bf
c}^{B(k)}_{gj}\cdot \frac{\partial {\bf n}}{\partial \theta})^2].
\eqno(4.18)
$$

By decomposing group velocity vector ${\bf c}^{B(k)}_{gj}$ in
system of orthonormalized vector ${\bf n}$, $\partial {\bf
n}/\partial \theta$, we have
$$
{\bf c}^{B(k)}_{gj}=c^{B(k)}_{gjn} {\bf n}+c^{B(k)}_{gj\theta}
\frac{\partial {\bf n}}{\partial \theta},
$$
with $c^{B(k)}_{gjn}=c^{B(k)}_{pj}$, $c^{B(k)}_{gj\theta}={\bf
c}^{B(k)}_{gj}\cdot (\partial {\bf n}/\partial \theta)$.

Therefore,
$$
|{\bf c}^{B(k)}_{gj}|^2=(c^{B(k)}_{pj})^2+\left( {\bf
c}^{B(k)}_{gj} \cdot \frac{\partial {\bf n}}{\partial \theta}
\right)^2,
$$
and as the result we obtain from (4.17), (4.18) the simple
expression
$$
q''=-k^{B(k)}_{pj} \frac{(c^{B(k)}_{gj})^2}{(c^{B(k)}_{pj})^4}.
\eqno(4.19)
$$

By substituting of (4.19) into (4.2), (4.1), (3.27) yields the
following formulae for far field asymptotic
$$
{\bf v}_d=\sum^2_{j=1}(\sum_k)' \sum^{N^{(k)}_j}_{m=1} {\bf
v}^{(k)}_{jm}, \quad \omega r \to \infty, \eqno(4.20)
$$
$$
{\bf v}^{(k)}_{jm} \approx (-1)^{k+1}\frac{if {\bf P}_j ({\bf
n}^{(k)}_{jm})\cdot {\bf l} \, c^{B(k)}_{pjm}} {2\sqrt{2\pi r}\rho
|{\bf c}^{B(k)}_{gjm}| c^A_{pjm} \sqrt{|k^{B(k)}_{pjm}|}} \, {\rm
e}^{-i (\omega r {\bf L}^{B(k)}_{jm} \cdot {\bf y} + \frac{\pi}4
\sigma^{B(k)}_{jm}) }, \eqno(4.21)
$$
where $N^{(k)}_j$ is the number of stationary point for separated
slowness curve, ${\bf L}^{B(k)}_{jm}={\bf L}^{B(k)}_{j} ({\bf
n}^{(k)}_{jm}) $, ${\bf c}^{B(k)}_{gjm}={\bf c}^{B(k)}_{gj} ({\bf
n}^{(k)}_{jm}) $, ${\bf c}^{B(k)}_{pjm}={\bf c}^{B(k)}_{pj} ({\bf
n}^{(k)}_{jm}) $, ${\bf c}^A_{pjm}={\bf c}^A_{pj} ({\bf
n}^{(k)}_{jm}) $, $k^{B(k)}_{pjm}=k^{B(k)}_{pj} ({\bf
n}^{(k)}_{jm}) $, $\sigma^{B(k)}_{jm} = {\rm sign}\,
k^{B(k)}_{pjm}$.

Because of $(-1)^k c^{B(k)}_{pjm}/c^A_{pjm} =\omega
/\Omega(\balpha^{(k)}_{jm})$, and in the stationary points from
(4.12), (4.13) ${\bf L}^{B(k)}_{jm} \cdot {\bf y} = |{\bf
c}^{B(k)}_{gjm}|^{-1}$, then we can rewrite formula (4.21) in the
form
$$
{\bf v}^{(k)}_{jm} \approx (-1)^{k+1}\frac{if\omega {\bf P}_j
({\bf n}^{(k)}_{jm})\cdot {\bf l}} {2\sqrt{2\pi r} \rho
\Omega(\balpha^{(k)}_{jm}) |{\bf c}^{B(k)}_{gjm}|
\sqrt{|k^{B(k)}_{pjm}|}} \, {\rm e}^{-i \left(\frac{\omega
r}{|{\bf c}^{B(k)}_{gjm}|} + \frac{\pi}4 \sigma^{B(k)}_{jm}\right)
}. \eqno(4.22)
$$

As we can see from (4.21), (4.22), in far zone the wave field is
separated in individual cylindrical waves $j=1,2$; $k=0,1$ or
$k=0$; $m=1,...,N^{(k)}_j$.

We note that the wave fields in the far zone have peculiarities in
the neighborhood of directions ${\bf n}^{(k)}_{jm} ({\bf x})$ for
which $k^{(k)}_{pjm} \approx 0$ for near located and for multiple
points. For such directions as in the problem $A$ [13] another
shapes of asymptotic decomposition than (4.20)--(4.22) are
required.

As it follows from (4.12), (4,13), the number of waves ${\bf
v}^{(k)}_{jm}$ in far field are determined by the number of
stationary value ${\bf n}^{(k)}_{jm} ({\bf x})$. As we marked
above, in the stationary value ${\bf n}^{(k)}_{jm} ({\bf x})$ the
group velocity vector ${\bf c}^{B(k)}_{gjm} ({\bf n}^{(k)}_{jm})$
is directed along the vector ${\bf x}$. Therefore, for fixed
direction ${\bf x}$ the number of waves is easy evaluated as the
number of intersection between the ray $0{\bf x}$ and the group
velocity curves. Depending on moving rate, the number of waves may
be essentially changed, and besides in the case of trans- and
superseismic motion the zones of a fast and slow waves
propagation, limited by Mach's cones, exist.

\includegraphics[bb = -35pt 0pt 780pt 340pt, scale=0.5]{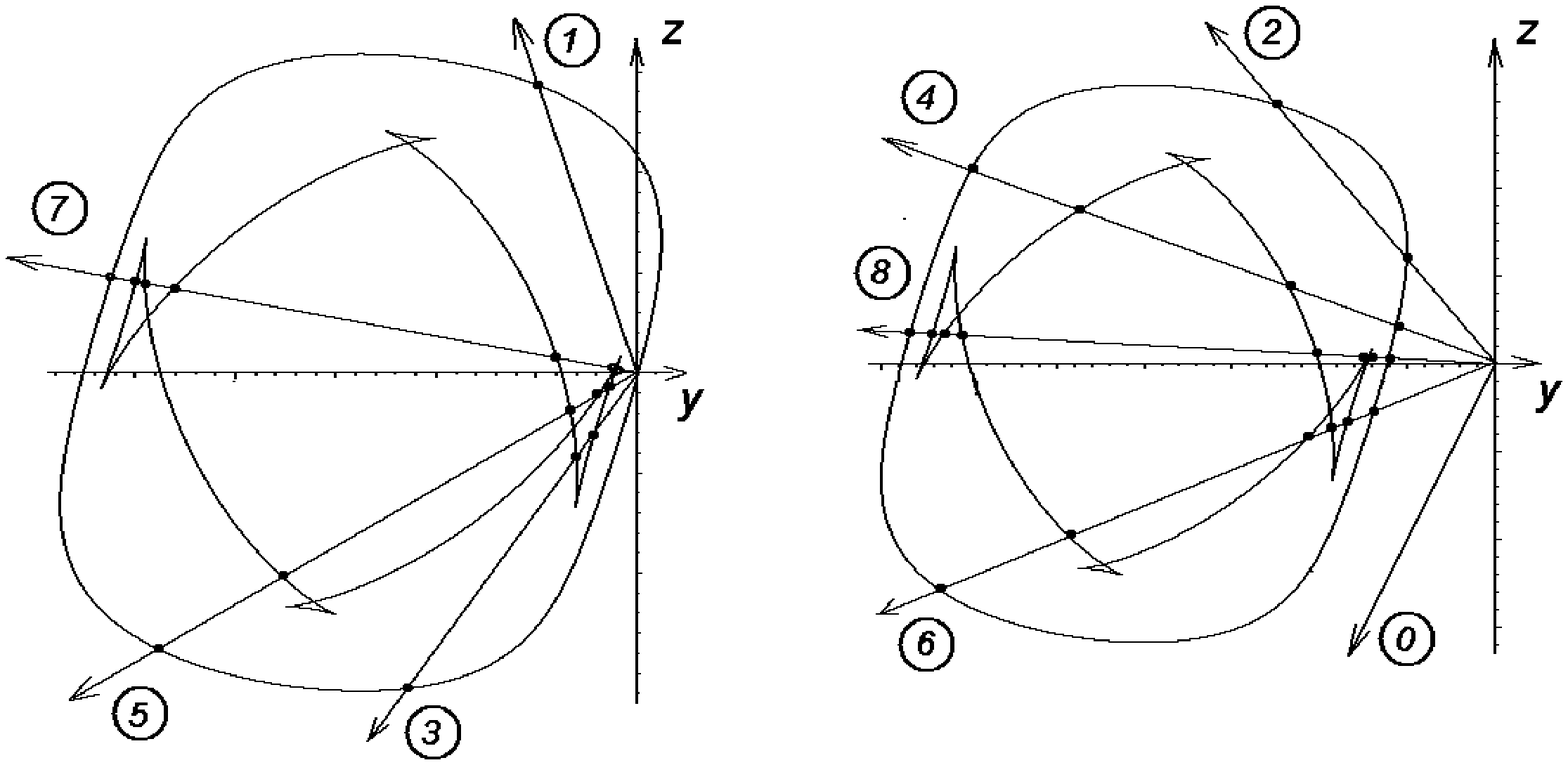}
\begin{center}
{Fig.~11. Group velocity curves, \qquad Fig.~12. Group velocity
curves, \qquad\\
\quad $\alpha$-quartz, $w_1=5500$, $w_2=0$. \qquad
$\alpha$-quartz, $w_1=8000$, $w_2=0$.}
\end{center}

The Figures ~11, 12 show typical pictures of group velocity curves
in planes which correspond to the work cartesian coordinate
system. Separately the selected directions ${\bf x}$ on the
Figs.~11, 12 are supplied with arrows and marked with numbers in
circles which denote the number of waves for these directions,
i.e. the number of intersection points of the curves $c^A_{gj}$
and $c^{B(k)}_{gj}$ with directions ${\bf x}$. On the figures the
points of intersection are bold-face points. Figs.~11, 12
illustrate, that the number of waves in different zones takes any
values from 0 to 8.

In pole coordinate system ($r$, $\tilde \theta$) the components of
group velocity vector in stationary points are given by:
$c^B_{gjmr} = |{\bf c}^B_{gjm}|$; $c^B_{gjm{\tilde \theta}} = 0$.
Hence, cylindrical waves (4.21), (4.22) in far field satisfy all
general conditions for cylindrical waves from [3]. Therefore, for
waves (4.21) or (4.22) the general energetic relations are correct
[3], and the group velocity vector ${\bf c}^B_{gjm}$ is equal to
energy transport velocity vector (ray velocity vector) for fixed
and moving observers.

We use general formulae [3] for the cylindrical waves average
energies $\langle E^{\xi x} \rangle$ and $\langle E^x \rangle$ in
moving coordinate system for fixed and moving observers
respectively
$$
\langle E^{\xi x} \rangle = \frac12 \Omega^2(\balpha) \rho {\bf v}
\cdot {\bf v}^*, \eqno(4.23)
$$
$$
\langle E^x \rangle = \frac12 \omega \Omega (\balpha) \rho {\bf v}
\cdot {\bf v}^*, \eqno(4.24)
$$
and formulae for average energy flux vector (Poynting's vector)
$$
\langle J^{\xi x (x)}_r \rangle = c^B_{gr} \langle E^{\xi x (x)}
\rangle, \eqno(4.25)
$$
where $\langle ... \rangle =\frac1T \int_0^T (...) \, dt $,
$T=2\pi /\omega$.

By substituting of (4.22) into (4.23)---(4.25) yields the
following expressions for energy flux of individual cylindrical
waves in far field
$$
\langle J^{\xi x (k)}_{jmr} \rangle = \frac{\omega^2 f^2 |{\bf
P}_j ({\bf n}^{(k)}_{jm})\cdot {\bf l}|^2} {16\pi \rho r |{\bf
c}^{B(k)}_{gjm}| |k^{B(k)}_{pjm}|}, \eqno(4.26)
$$
$$
\langle J^{x (k)}_{jmr} \rangle =
\frac{\omega}{\Omega(\balpha^{(k)}_{jm})} \langle J^{\xi x
(k)}_{jmr} \rangle = (-1)^k \frac{c^{B(k)}_{pjm}}{c^A_{pjm}}
\langle J^{\xi x (k)}_{jmr} \rangle. \eqno(4.27)
$$

For superseismic moving rate $\Omega(\balpha^{1}_{jm})<0$ for
$k=1$, and therefore $\langle J^{x 1}_{jmr} \rangle <0$, i.e. the
slow waves transfer the negative energy, measured by moving
observer. But this property is common for slow waves in problem
$B$ with superseismic moving sources.

The Figures ~13--22 show the average energy flux curves $\langle
J^{\xi x (k)}_r \rangle$ for fused silica and $\alpha$-quartz
($SiO_2$). The normalizing factor $\zeta$ for the values of
average energy flux is equal to $\omega^2 f^2/(16\pi \rho r)$.

The energy flux curves have more complicated forms, then slowness
and group velocity curves. Really, according to (4.26) the values
of energy flux depend on the values of group velocity, curvature
of slowness curve and form the scalar product between polarization
vector ${\bf p}^{(k)}_{jm}={\bf p}_j ({\bf n}^{(k)}_{jm})$ and the
unit vector of source direction ${\bf l}$ (by (3.8) $|{\bf P}_j
({\bf n}^{(k)}_{jm})\cdot {\bf l}|=|{\bf p}_j ({\bf
n}^{(k)}_{jm})|$). For some angle $\theta$ the polarization
vectors can be orthogonal to unit vector of source direction ${\bf
l}$, because quasi-longitudinal and quasi-shear polarization
vectors are perpendicular, and if the angle $\theta$ change by
$2\pi$, the polarization vectors also rotate by $2\pi$. In these
cases the energy flux vector is equal to zero. On the other hand
the energy flux tends to infinity, when then curvature of slowness
curve tends to zero. This exists, when slowness curves stretch on
infinity and in the points of inflection. As a result, taking into
account the possibility of existence of zones with different
number of spread waves, the energy flux curves are more
complicated and complex for analysis.

The Figures ~13--18 show the results of energy flux calculations
for isotropic material fused silica. The values of source motion
are equal to corresponding values for Figs.~1-6 (for Figs.~13, 14
$w_1=w_2=0$, for Figs.~15, 16 $w_1=2.6\cdot 10^3$ m/s, for
Figs.~17, 18 $w_1=7.5\cdot 10^3$ m/s). The consistent pair of the
figures correspond to directions of unit vector of source along
motion direction ($l_1=1$, $l_2=0$ for Figs.~13, 15, 17), and
perpendicular to motion direction ($l_1=0$, $l_2=1$ for Figs.~14,
16, 18). As it was used above, the curves marked subscript "1"
correspond to the results for quasi-shear waves and the curves
marked subscript "2" correspond to the results for
quasi-longitudinal waves.

\includegraphics[bb = -75pt 20pt 740pt 270pt, scale=0.5]{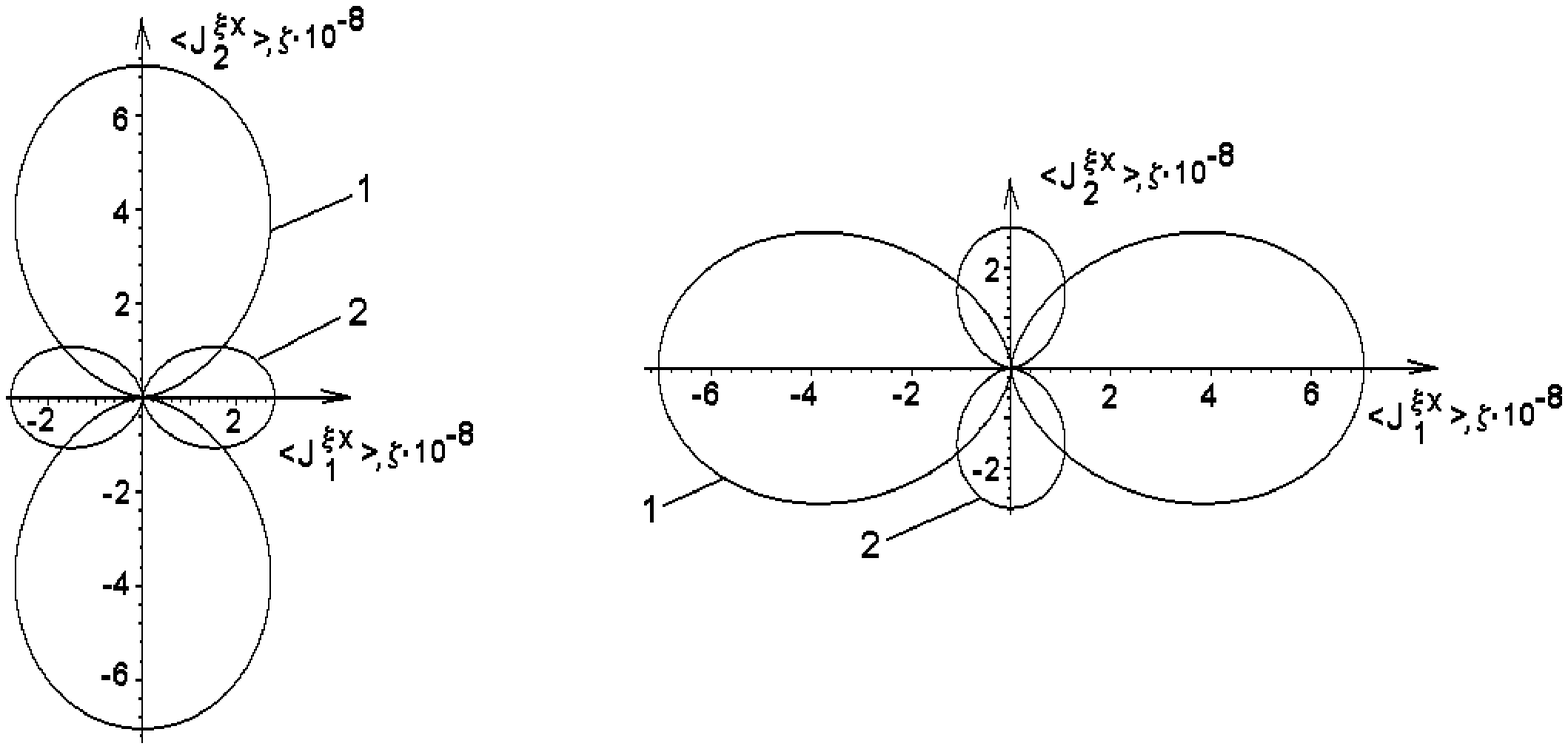}
\begin{center}
{Fig.~13. Energy flux curves, \qquad  \qquad Fig.~14. Energy flux curves, \qquad\\
\quad fused silica, $w_1=w_2=0$, $l_1=1$. \quad fused silica,
$w_1=w_2=0$, $l_2=1$.}
\end{center}

\includegraphics[bb = -35pt 15pt 780pt 315pt, scale=0.5]{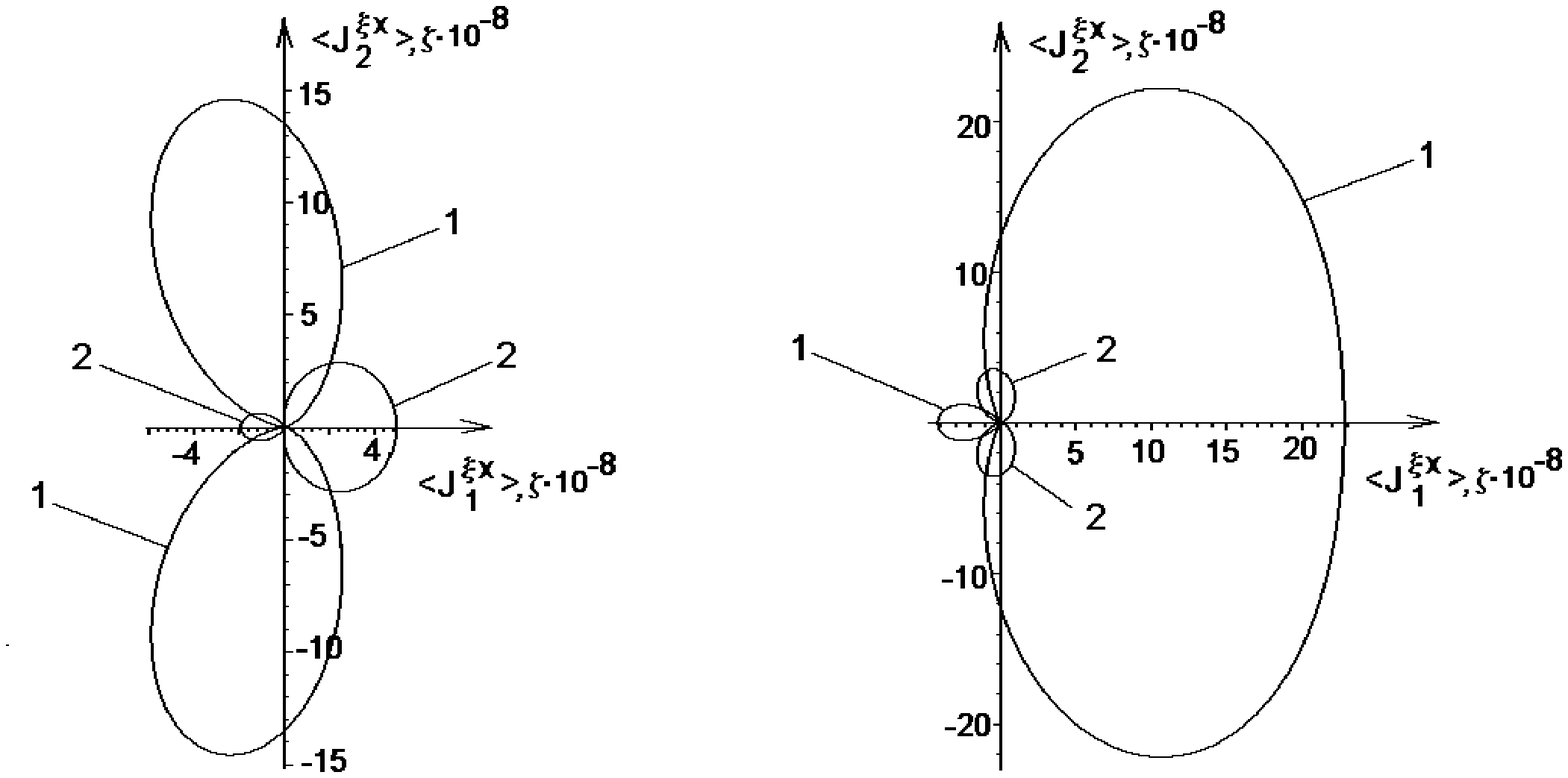}
\begin{center}
{Fig.~15. Energy flux curves, \qquad  \qquad Fig.~16. Energy flux curves, \qquad\\
\quad fused silica, $w_1=2600$, $w_2=0$, $l_1=1$. \quad fused
silica, $w_1=2600$, $w_2=0$, $l_2=1$.}
\end{center}

\includegraphics[bb = -35pt 0pt 780pt 310pt, scale=0.5]{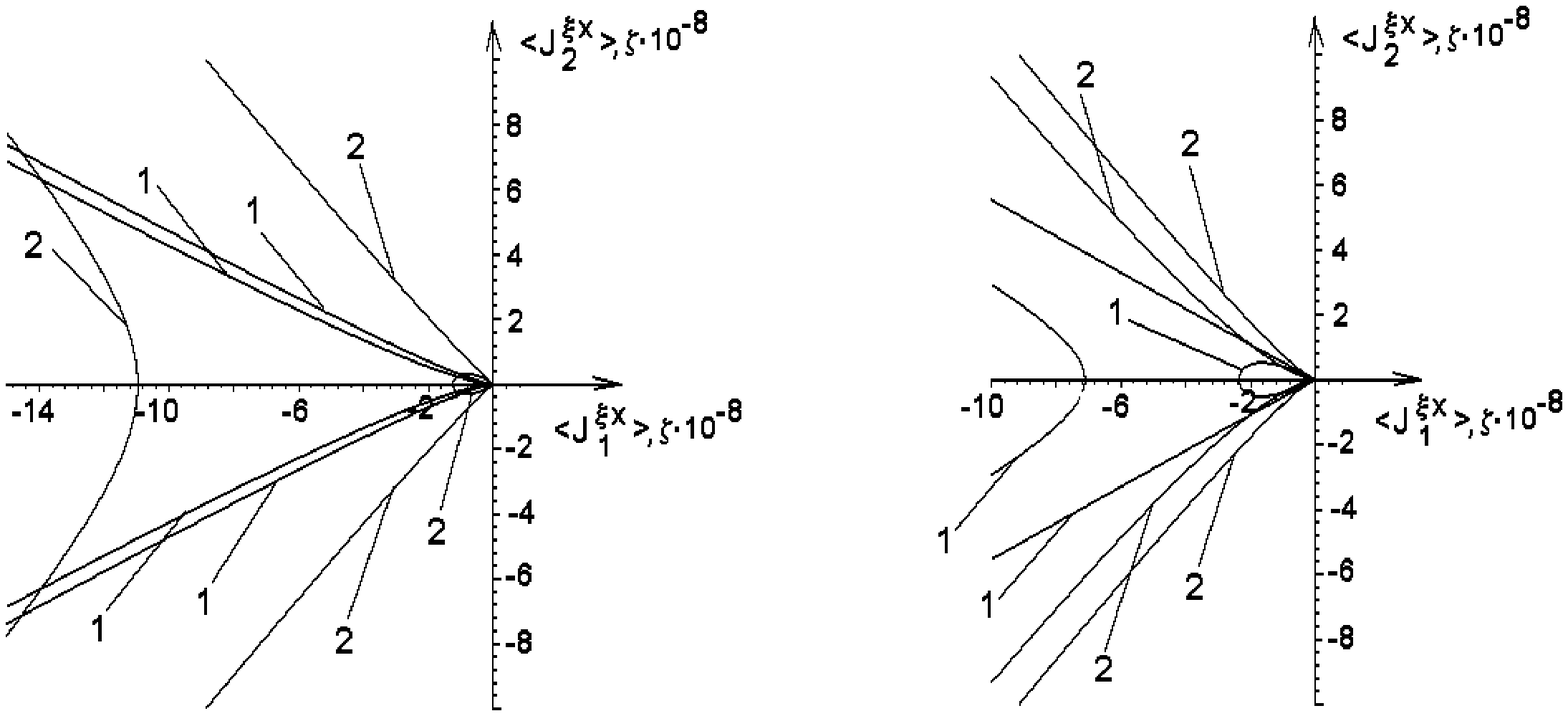}
\begin{center}
{Fig.~17. Energy flux curves, \qquad  \qquad Fig.~18. Energy flux curves, \qquad\\
\quad fused silica, $w_1=7500$, $w_2=0$, $l_1=1$. \quad fused
silica, $w_1=7500$, $w_2=0$, $l_2=1$.}
\end{center}

From the Figs.~13, 14 it is obvious, that in the problem $A$ for
isotropic material fused silica there is the maximum of energy
flux for quasi-share wave in direction, which is perpendicular to
direction of unit vector of source. All curves of energy source in
the problem $A$ are symmetrical to the origin of coordinate
system. For the problem $B$ with subseismic motion (Figs.~15, 16)
the curves of energy source change significantly and become not
symmetrical. For superseismic motion (Figs.~17, 18) the energy
flux curves are not limited, because not far from boundaries of
wave propagation (Mach's cones) the curvature of slowness curves
tends to 0. As for the case of slowness curves each couple of
energy flux curves with superseismic motion corresponds to fast
and slow waves.

The figures ~19--22 illustrate the pictures of energy flux
behavior for anisotropic material ($\alpha$-quartz).  For the
Figs.~19, 20 we have the case of problem $A$ with ${\bf w}=0$
(analogically Figs.~7, 8), and for the Figs.~21, 22 we have the
case of problem $B$ with superseismic motion (analogically
Figs.~9, 10). In this cases we obtain exceptionally complicated
behavior of energy flux vectors. (For obviousness the subscript
"1", which corresponds to quasi-shear waves is absent in
Figs.~19--22.)

\includegraphics[bb = -35pt 0pt 780pt 330pt, scale=0.5]{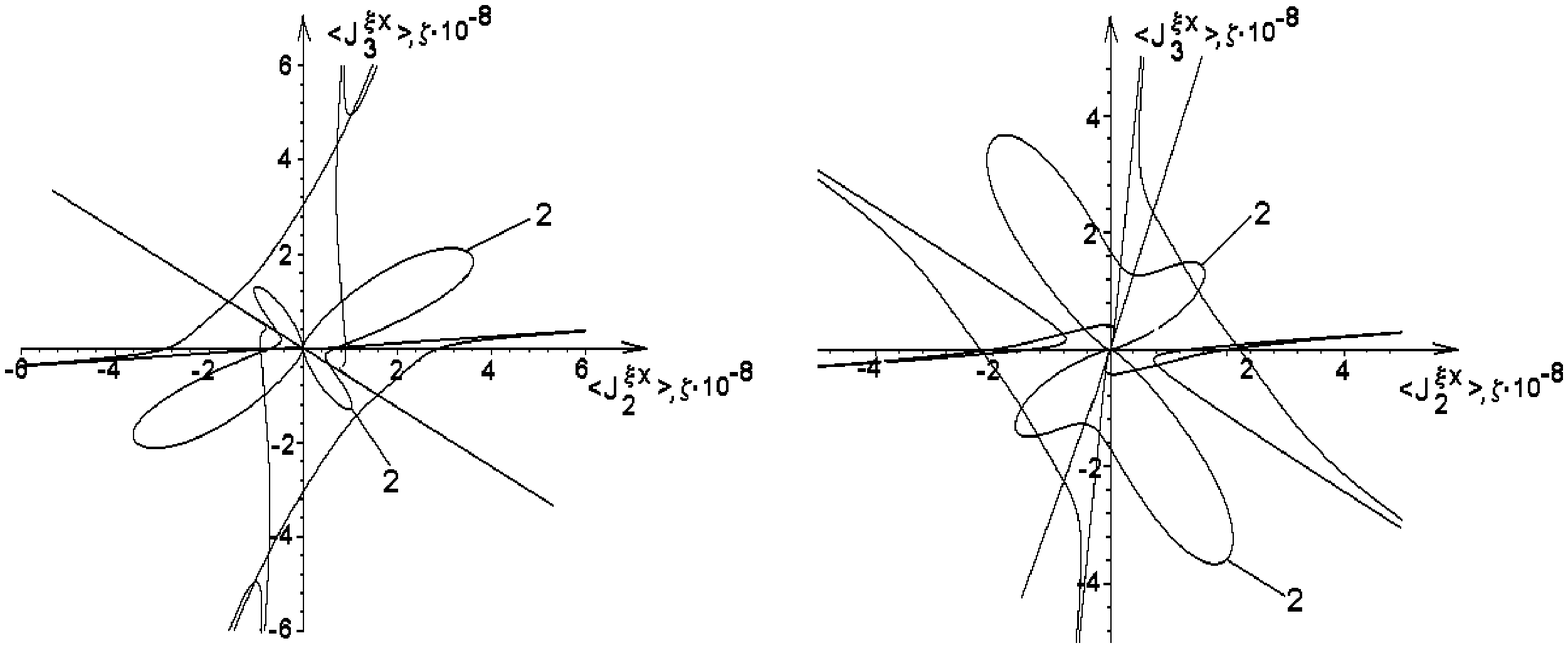}
\begin{center}
{Fig.~19. Energy flux curves, \qquad  \qquad Fig.~20. Energy flux curves, \qquad\\
\quad $\alpha$-quartz, $w_1=w_2=0$, $l_1=1$. \quad
$\alpha$-quartz, $w_1=w_2=0$, $l_2=1$.}
\end{center}

\includegraphics[bb = -35pt 0pt 780pt 320pt, scale=0.5]{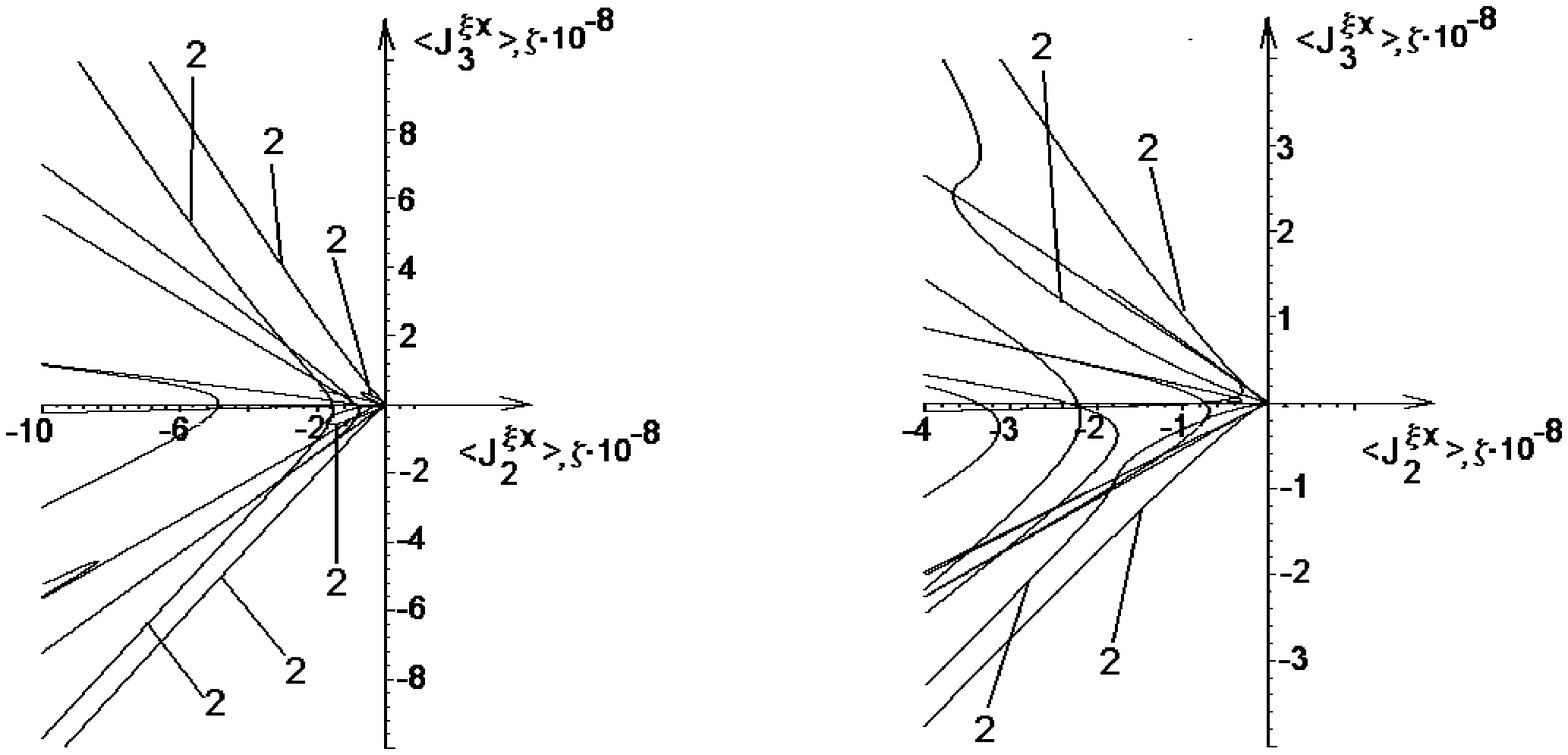}
\begin{center}
{Fig.~21. Energy flux curves, \qquad  \qquad Fig.~22. Energy flux curves, \qquad\\
\quad $\alpha$-quartz, $w_1=8000$, $w_2=0$, $l_1=1$. \quad
$\alpha$-quartz, $w_1=8000$, $w_2=0$, $l_2=1$.}
\end{center}

In the conclusion we announce similar paper [14], where additional
pictures of phase velocity are presented, but the results of
energy flux calculation are absent.

%{\bf Acknowledgements}

%The paper has been supported by Italian Ministry of University
%(M.U.R.S.T.) through its national and local (60\%) projects.

%\newpage

\bigskip

% BELOW SHOULD BE YOUR ADDRESSES
\address{Gerardo Iovane, D.I.I.M.A., University of Salerno, 84084 Fisciano (SA), Italy. Tel: +39 089
96 42 68. Fax: +39 089 96 41 91.}
\address{Andrei V.~Nasedkin, Faculty of Mechanics and Mathematics,  Rostov
State University, Zorge 5, Rostov-on-Don 344090, Russia. Tel: +007
(8632) 43 47 11. Fax: +007 (8632) 64 52 55.}
\address{Michele Ciarletta, D.I.I.M.A., University of Salerno, 84084 Fisciano (SA), Italy. Tel: 39 089
96 42 51. Fax: 39 089 96 41 91.}

\end{document}